\documentclass[aps,prl,twocolumn,superscriptaddress,showpacs]{revtex4}

\usepackage{graphicx}
\usepackage{dcolumn}
\usepackage{bm}

\def\kdp{{\bf k}$\cdot${\bf p}}

\newcounter{Alist}

\begin{document}

\preprint{APS/123-QED}

\title{Ab-initio Prediction of Conduction Band Spin Splitting in Zincblende
Semiconductors}

\author{A. N. Chantis, Mark van Schilfgaarde and Takao Kotani}
\affiliation{Arizona State University, Tempe, Arizona, 85284, USA}
\date{\today}

\begin{abstract}

We use a recently developed self-consistent $GW$ approximation to present
systematic \emph{ab initio} calculations of the conduction band spin
splitting in III-V and II-V zincblende semiconductors.  The spin orbit
interaction is taken into account as a perturbation to the scalar
relativistic hamiltonian.  These are the first calculations of conduction
band spin splittings based on a quasiparticle approach; and because the
self-consistent $GW$ scheme accurately reproduces the relevant band
parameters, it is expected to be a reliable predictor of spin splittings.
The results are compared to the few available experimental data and a
previous calculation based on a model one-particle potential.  We also
briefly address the widely used \kdp\ parameterization in the context
of these results.


\end{abstract}

\pacs{71.70.-d, 71.70.Ej, 71.15.-m, 71.15.Qe ,71.15.Mb )}
\maketitle

In this letter, we apply a recently developed all-electron quasiparticle
(QP) self-consistent $GW$ method (QPsc$GW$) \cite{Mark2,Mark3,Mark4} to
determine the spin splitting of conduction bands in III-V and II-VI
zincblende semiconductors as a consequence of the spin-orbit coupling.
This quantity is emerging as a parameter of great importance in spintronics
applications, in part because it determines spin lifetimes, but also
because it may be possible to exploit the splitting to induce spin current
without an external magnetic field through the Rashba effect.

Information about this important quantity is very sparse.  On the
theoretical side, precise QP levels for the occupied and first few
unoccupied states are required.  We show here that QPsc$GW$ accurately
reproduces measured fundamental and higher-lying gaps in all the
semiconductors studied, as well as other key parameters (effective mass,
$d$ band position, etc.), and is therefore a suitable vehicle for accurate
determination of the splitting.


Owing to a lack of inversion symmetry, the conduction band of a zincblende
semiconductor is split by spin-orbit coupling.  Dresselhaus \cite{dress}
showed that near the minimum point $\Gamma_{6}^{c}$, the splitting can be
described by the operator
$H_{\rm{}D}=\frac{1}{2}\vec{\sigma}\cdot\vec{B}_{\rm{}eff}$, where
$\vec{\sigma}$ are the Pauli matrices and $B_{\rm eff}^{i}=2\gamma k_{i}
(k^{2}_{i+1}-k^{2}_{i+2})$, $i=\{x,y,z\}$.  $H_{\rm{D}}$ is written in this
form to suggest that its effect is equivalent to the action of a
{\bf{k}}-dependent effective magnetic field $\vec{B}_{{\rm{eff}}}$, which
is anisotropic in ${\bf{k}}$ and is proportional to $k^3$.  $\gamma$ is a
constant that depends on the bulk properties of the material.  Provided the
Larmor precession frequency under $\vec{B}_{{\rm eff}}({\bf{k}})$ is less
than the inverse of momentum relaxation time, small random changes in
${\bf{k}}$ will cause the polarization of an injected current to gradually
diminish.  This is the spin scattering mechanism of Dyakonov and Perel (DP)
\cite{Dyakonov}; and it is widely accepted that for a wide range of
temperatures and moderate concentration of impurities, DP is the dominant
spin relaxation mechanism in the conduction band of III-V and II-VI
semiconductors.  The DP spin scattering rate can be estimated theoretically
once the value of $\gamma$ is known \cite{Dyakonov}.

In low dimensional structures the splitting scales in proportion to $k$,
owing to confinement effects, which causes the DP mechanism to increase
significantly.  Moreover asymmetry in the leads or structure of a device results in an
additional spin splitting effect, proposed first by Rashba and Bychkov
\cite{Rashba}.  This contribution from this ``structural inversion
asymmetry'' depends on the boundary conditions: it can interfere
constructively or destructively with the Dresselhaus term depending on
device geometry and {\bf k} vector.  Based on this, a few authors
\cite{Dyakonov2,Averkiev,Xavier} proposed that growth of quantum wells in certain
crystallographic directions can result in long spin coherence times, if the
ratio of Dresselhaus to Rashba terms is close to unity.  Measurements of
these two contributions suggest that they can be comparable
\cite{Knap,Jusserand}.  Also giant spin relaxation anisotropy can occur in
strained bulk zincblende semiconductors owing to destructive interference
between the Dresselhaus and strain-induced spin splittings \cite{Averkiev}.

To evaluate $\gamma$ theoretically, it is essential to use an \emph{ab
initio} approach rather than a parameterized method such as the \kdp\
theory because the latter includes many external parameters whose values
can only be reliably determined through fitting to an {\em ab initio}
approach.  However, the \emph{ab initio} approach based on the usual local
density approximation (LDA) is no better.  This is because $\gamma$ depends
strongly on key features of the band structure, such as effective mass,
that the LDA cannot reliably predict. 
As is well known, it is difficult to choose an appropriate empirical
adjustment to the LDA potential so that the energy band structure of the
lowest few conduction bands is reliable.  This is particularly true in more
complex cases such as quantum wells where experimental data is not
available.




\begin{table*}
\caption{\label{tab:table1} Important band parameters for the III-V
  semiconductors.  $E_{0}$ and $E^{\prime}_{0}$ are the energies of the
  first two conduction bands at the $\Gamma$-point, and are defined in the
  text; $E_{g}$ is the fundamental gap when it differs from
  $E_{0}$. $m^{\Gamma}_{c}/m$ is the conduction band effective mass at
  $\Gamma$.  Energies are in eV; $\gamma$ is in eV$\cdot\rm{}\AA^{3}$.
  Quantities in rectangular brackets were computed from Eq.~2 with $\alpha$
  chosen to reproduce the experimental fundamental gap $E_g$.  Quantities
  in parenthesis and curly brackets correspond respectively to experimental
  data, and calculated data from Ref.~\onlinecite{Cardona}.  Spin orbit
  splittings $\Delta_{\rm{}SO}$ and $\Delta^{\prime}_{\rm{}SO}$ do not
  change significantly whether the QPsc$GW$, the scaled QPsc$GW$ or the LDA
  potential is used.}
\begin{ruledtabular}


\begin{tabular}{lccccccccc}
                               &   AlP   &    AlAs   &    AlSb     &    GaP   &     GaAs  &   GaSb  &   InP    &   InAs     &InSb     \\
\hline
$E_{g}$                        & 2.61   &   2.25   &   1.75     &   2.33  &           &         &          &            &         \\
(expt)                          & (2.51)\footnotemark[1]&   (2.23)\footnotemark[1]&   (1.69)\footnotemark[1]&   (2.35)\footnotemark[1]   &           &         &          &            &         \\
\hline
$E_{0}$                        & 4.52   &   3.33   &   2.66     &   3.00  &   1.80   &  1.16  &  1.56    &  0.68     &  0.54  \\
(expt)                          &(3.63)\footnotemark[1]&  (3.13)\footnotemark[1]&  (2.38)\footnotemark[1]&  (2.90)\footnotemark[3] &  (1.52)\footnotemark[1]   & (0.82)\footnotemark[1]  & (1.42)\footnotemark[1]    & (0.42)\footnotemark[6]     & (0.24)\footnotemark[1]  \\
\hline
$E^{\prime}_{0}-E_{0}$        & 1.46   &   2.13   &   1.22     &   1.98  &   2.81   &  2.10  &  3.32   &  3.10     &  2.69  \\
{}[scaled $\Sigma$]           &        &          &            &         &  [2.89]  & [2.26] & [3.34]  & [3.78]    & [2.82] \\
(expt)                         &        &          &            &         &  (3.08)\footnotemark[1]
                                                                                    & (2.37)\footnotemark[1]
										             & (3.38)\footnotemark[1]
													  &        & (2.91)\footnotemark[1]  \\
\hline
$\Delta_{{\rm SO}}$            & 0.060   &   0.294   &   0.664     &   0.096  &   0.336   &  0.703  &  0.12    &  0.359     &  0.733  \\
(expt)                          &
                                         &  (0.300)\footnotemark[1]
                                                     &  (0.673)\footnotemark[1]
                                                                   &          &  (0.341)\footnotemark[1]
              								                  & (0.756)\footnotemark[1]
											            & (0.108)\footnotemark[1]
													      &  (0.371)\footnotemark[7]&  (0.750)\footnotemark[8]  \\
\hline
$\Delta^{\prime}_{{\rm SO}}$ & 0.027   &  0.031   &  0.053     &  0.158  &   0.174   &  0.196  &  0.423   &  0.429     &  0.389  \\
(expt)                        &         &          &            &         &           & (0.213)\footnotemark[1]
                                                  					       & (0.070)\footnotemark[1]
													  &            & (0.392)\footnotemark[1]  \\
\hline
$\Delta^{-}$                 & -0.10   &  -0.13    &   -0.32     &  +0.12 &  -0.12 &  -0.32  &  +0.21   &  +0.22     &  -0.29   \\
{}\{Ref.~\onlinecite{Cardona}\} &      &           &             &             & \{-0.11\}& \{-0.32\}& \{+0.23\}&         & \{-0.24\} \\
\hline
$m^{\Gamma}_{c}/m$           & 0.186   &   0.131   &   0.117     &   0.130  &   0.076   &  0.055  &  0.084   &  0.036     &  0.030  \\
{}[scaled $\Sigma$]          &         &           &             &          &  [0.069]  & [0.043] & [0.081]  & [0.026]    & [0.016] \\
(expt)                        &         &           &             &          &  (0.067)\footnotemark[1]
										    & (0.0412)\footnotemark[1]
											      & (0.0765)\footnotemark[1]
													 & (0.0231)\footnotemark[1]
														      &(0.014)\footnotemark[1]  \\
\hline
$\gamma$                 &+0.08    &  +3.4   &  +30.2     & -2.3   &  +6.4   & +81.8 &	-14.1  &  -27.8    & +79.4 \\
{}[scaled $\Sigma$]      &         &          &           &         & [+8.5]  &[+119.3]&[-15.7] &[-47.5]   &[+209.6]\\
(expt)                    &         &          &           &         &  (25.5)\footnotemark[9]
									      & (186.2)\footnotemark[10]
											      &(7.5-9.9)\footnotemark[10]&
													 &(+225$\pm$12)\footnotemark[11]     \\
{}\{Ref.~\onlinecite{Cardona}\}   &&           &          &         & \{+14.9\}&\{+108.8\}    &{-8.9}       &            & \{+217.6\}    \\
\end{tabular}
\end{ruledtabular}
\footnotetext[1]{From Ref.~\cite{madelung}}
\footnotetext[3]{From Ref.~\cite{Nelson}}
\footnotetext[6]{From Ref.~\cite{Varfolomeeve}}
\footnotetext[7]{From Ref.~\cite{Landolt}}
\footnotetext[8]{From Ref.~\cite{Jung}}
\footnotetext[9]{From Ref.~\cite{Marus}}
\footnotetext[10]{From Ref.~\cite{Gorel}}
\footnotetext[11]{From Ref.~\cite{Chen}.}
\end{table*}

QPsc$GW$ is a method to determine a (nonlocal but static) hermitian
one-particle hamiltonian $H_0$ in a self-consistent way \cite{Mark2,Mark3}.
With the usual $GW$ approximation, we can calculate the self-energy
$\Sigma(\omega)$ as a functional of the starting trial $H_0$.
Here $\Sigma(\omega)$ can be expanded in the basis of the
eigenfunctions $\{\psi_{{\bf{k}}n}({\bf r}) \}$ of $H_0$, where ${\bf{k}} $ is
the wave vector and $n$ is the band index.
Then we define the static self-energy as
\begin{eqnarray}
\widetilde{\Sigma}_{{\bf{k}}nn'} = {\rm Re}
\langle \psi_{{\bf{k}}n}| \frac{\Sigma(\epsilon_{{\bf{k}}n}) +
\Sigma(\epsilon_{{\bf{k}}n'})}{2} | \psi_{{\bf{k}}n'}\rangle
\end{eqnarray}
where, $\epsilon_{{\bf{k}}n}$ denote eigenvalues, and ${\rm Re}$ means to
take the hermitian part.
By using this $\widetilde{\Sigma}_{{\bf{k}}nn'}$ instead of the usual LDA
exchange correlation potential $V^{\rm LDA}_{\rm xc}$, we can iterate
to self-consistency, i.e. the  $\widetilde{\Sigma}_{{\bf{k}}nn'}$
generated by $H_0$ is identical to the $\widetilde{\Sigma}_{{\bf{k}}nn'}$
that enters into $H_0$.
We explicitly include core contributions to $\Sigma$.


With the addition of local orbitals, we include in the MTO basis all atomic-like
states that are within $\sim\pm$2~Ry of the Fermi energy: thus
both the $3d$ and $4d$ levels are included for GaAs, and so on.
As QPsc$GW$ gives the self-consistent $H_0$ at the scalar relativistic
level, we add the spin-orbit operator
$H_{{\rm SO}}=\vec{L}\cdot\vec{S}/2c^2$ to $H_0$ as a perturbation (it is
not included in the self-consistency cycle).
Thus the band structure is calculated from a one-body Hamiltonian $H_0 + H_{\rm SO}$.
We also considered a `scaled $\Sigma$' with
\begin{eqnarray}
H_\alpha +  H_{\rm SO} = H_{\rm LDA} + (1-\alpha)(\widetilde{\Sigma}-
V^{\rm LDA}_{\rm xc}) + H_{\rm SO}.
\end{eqnarray}
(Note that $H_{\rm LDA}$ and $V^{\rm LDA}_{\rm xc}$ are also determined
from the self-consistent density given by $H_0$).  Tables I and II present
QPsc$GW$ results ($\alpha=0$) for III-V and II-VI compounds.  We also
present results in a few cases where $\alpha$ is chosen to reproduce the
experimental fundamental gap $E_{g}$ at 0K, which we do to address the
strong sensitivity of $\gamma$ to $E_{g}$.  The error in $E_{g}$ from
QPsc$GW$ theory, while small, is comparable to $E_{g}$ itself for small-gap
systems InAs and InSb, resulting in large errors in the effective mass
$m^{\Gamma}_{c}$.  Evidently once $(\widetilde{\Sigma}-V^{\rm LDA}_{\rm
xc})$ is scaled so $E_g$ is reproduced, $m^{\Gamma}_{c}$ falls in good
agreement with the experimental mass.  The magnitude of $\gamma$ depends on
the hybridization between the lower conduction band and the $p$-like
conduction and valence bands.  The hybridization depends on the position of
the first two conduction band levels
$E_{0}=E(\Gamma^{c}_{6})-E(\Gamma^{v}_{8})$ and
$E^{\prime}_{0}=E(\Gamma^{c}_{7})-E(\Gamma^{v}_{8})$.  (In the absence of
spin orbit coupling, $\Gamma^{c}_{6}$ has $\Gamma_{1}$ symmetry;
$\Gamma^{v}_{8}$,$\Gamma^{v}_{7}$ and $\Gamma^{c}_{8}$,$\Gamma^{c}_{7}$ are
the valence and conduction levels with $\Gamma_{15}$ symmetry.) and is
reflected in the magnitude of effective mass $m^{\Gamma}_{c}$.  Also
important is the strength of spin-orbit interaction at anion and cation
sites, reflected in the spin orbit splittings $\Delta_{{\rm SO}}$ and
$\Delta^{\prime}_{{\rm SO}}$ of $\Gamma^{v}_{15}$ and $\Gamma^{c}_{15}$.
With the exceptions of $\Delta^{\prime}_{{\rm SO}}$ in InP and $E_{0}$ in
AlP, all parameters are in good agreement with available experimental data.
The experimental $\Delta^{\prime}_{{\rm SO}}$ seems low in InP and
inconsistent with the trends described below.  The discrepancy in $E_{0}$
for AlP is also curious as it falls well outside the error usually expected
from QPsc$GW$\cite{Mark3}; also its better known \emph{fundamental} gap
agrees well with QPsc$GW$.  It is likely that $E_{0}$ was misidentified in
the (30 year old) experiment.


As the valence band has mostly anion character and the conduction band
mostly cation character, one would expect that $\Delta_{{\rm SO}}$ is
controlled mainly by the anion, while $\Delta^{\prime}_{{\rm SO}}$ is
controlled mainly by the cation.  Comparing compounds with the same cation
but different anions, or vice-versa, this trend becomes evident.  For
example, for GaP, GaAs and GaSb $\Delta^{\prime}_{{\rm SO}}$ is 0.158,
0.174, and 0.196 eV respectively; while on the other side for AlAs, GaAs
and InAs $\Delta^{\prime}_{{\rm SO}}$ is 0.031, 0.174 and 0.429 eV.  This
trend is found in all materials presented in Tables I and II.
\begin{figure}[b]
\includegraphics[]{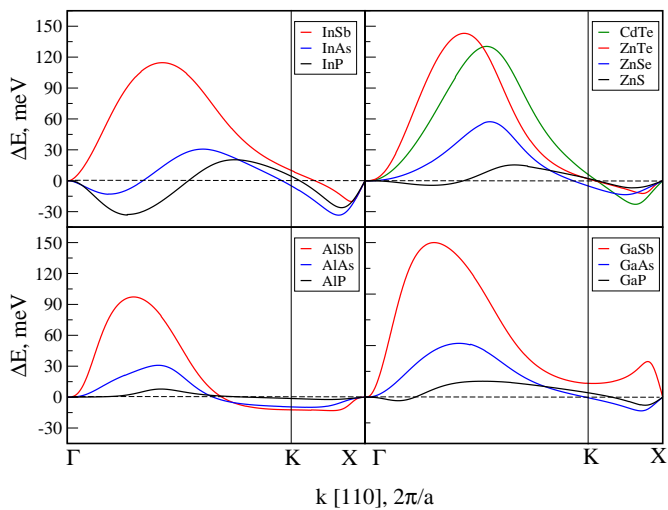}
\caption{\label{fig:epsart} Conduction band spin splitting in [110]
direction.  GaAs, GaSb, InP, InAs and InSb are plotted for the scaled $\Sigma$.}
\end{figure}

Fig.~\ref{fig:epsart} shows the conduction band spin splitting
versus wave number for the entire [110] line.  For sufficiently
small $k$ it varies as $k^3$ as expected from the
\kdp\  analysis; $\gamma$ was obtained by
fitting a polynomial to the calculated splitting at $k$ points
near $\Gamma$.  The sign of $\gamma$ was determined from the
eigenvectors and is defined according to the most accepted
convention ($\Delta E = E(\Sigma_{4})-E(\Sigma_{3})$).

The splitting is rather complex, arising from several factors such as how
the $\Gamma^v_{15}$ and $\Gamma^c_{15}$ states admix with the
conduction state as $k$ moves away from $\Gamma$.  Still, inspection of
Table~1 and Fig.~1 reveals some trends:

\setcounter{Alist}{0}

\begin{list}{({\it \roman{Alist}})\,}{\leftmargin 0pt \itemindent 18pt \usecounter{Alist}\addtocounter{Alist}{0}}

\item $\gamma$ scales approximately in proportion to $1/E_0$, at least in
      the narrow-gap case, as can be seen by comparing scaled
      ($H_{\alpha}+H_{\rm SO}$) and unscaled ($H_{0}+H_{\rm SO}$) results
      for InAs and InSb.  This is expected by the \kdp\ 
      theory.  The scaling only slightly perturbs the potential; the band
      structure stays approximately constant except for small shifts in the
      lowest conduction band that substantially affect the gap and the
      conduction band shape near the $\Gamma$-point (and the valence band
      that couples to this band).

\item $\gamma$ tends to increase as the spin-orbit coupling parameter
      $\Delta^{-}=3\left<(\frac{3}{2}\frac{3}{2})_{{\rm v}}|H_{{\rm
      SO}}|(\frac{3}{2}\frac{3}{2})_{{\rm c}}\right>$ becomes more
      negative.  $\Delta^{-}$ becomes more negative with increasing anion
      mass, while the reverse is true when cation mass increases.

\end{list}

When anion mass increases, these tendencies add constructively and $\gamma$
increases monotonically.  When cation mass increases, these tendencies
interfere with each other, and the behavior is more complex.

Table I compares our calculations of $\gamma$ with the work of Cardona et
al \cite{Cardona}.  Those authors added empirical pseudo-Darwin potential
shifts to make their band structure match the observed bandgap.  These
shifts result in a reasonable and qualitatively correct band structure,
though band parameters have varying degrees of accuracy
(e.g. $m^{\Gamma}_{c}$ deviated from experiment as much as
35\%\cite{Cardona}).  The splittings we obtain are in qualitative, and
often quantitative agreement with Cardona's calculations; e.g. $\Delta E$
changes sign at approximately the same $k$-points in GaP and GaAs.

What limited experimental data that exists is also shown in Table I.  For
GaAs \cite{Marus}, GaSb \cite{Gorel}, InP \cite{Gorel}, $|\gamma|$ was
inferred from an estimation of the spin relaxation rate ascribed to the DP
spin relaxation mechanism.  This method does not resolve the sign of
$\gamma$.  For InSb $\gamma$ was also extracted from
electric-dipole--magnetic-dipole interference at far infrared frequencies
\cite{Chen}, which is a more direct measure of the splitting and resolves
the sign.  This is probably the most accurate measurement of $\gamma$ in
any III-V semiconductor.  The measurements of $\gamma$ in
Refs.~\cite{Marus, Gorel} are rather indirect and are based on extracting
simultaneously and accurately spin and momentum relaxation times.  Also
there can be some uncertainty related to the type of momentum scattering
mechanism.

Some care must be taken to determine $\gamma$ in InSb.  Owing to the small
gap, the splitting begins to deviate from a $k^3$ dependence at very small
$k$ \cite{Pikus}.  For the electron densities used in that work
($N_{e}$=1.6$\times{}10^{14}$ to $4.7\times10^{15}$ cm$^{-3}$) the Fermi
wave number $k_{F}$ ranged between $0.9 \times 10^{-3}$ and
$2.7\times{}10^{-3}$ a.u.$^{-1}$.  At this $k$, we find a small deviation
from the $k^3$ dependence ($\sim$6\%).  However at larger doping the
renormalization is significant:  at $k=10\times{}10^{-3}$ a.u.$^{-1}$ the
splitting is reduced by 40\% (i.e. $\Delta E/\gamma k^3=0.6$) in InSb,
by 20\% in InAs, but is negligible in the remaining semiconductors.

\begin{table}[t]
\caption{\label{tab:table2} Important band parameters for the II-VI
  semiconductors.  Symbols are the same as in Table I.  Experimental
  data are taken from Ref.~\onlinecite{madelung} unless noted otherwise.}
\begin{ruledtabular}

\begin{tabular}{lcccc}
                               &   ZnS   &    ZnSe  &    ZnTe      &  CdTe  \\

\hline
$E_{0}$         &   4.00 &    2.94  &   2.45     &  1.84 \\
(expt)          &  (3.78)&   (2.82) &  (2.394)   & (1.606) \\
\hline
$E^{\prime}_{0}-E_{0}$&  4.59  &   4.52   &   2.91     &  3.57 \\
\hline
$\Delta_{{\rm SO}}$           &  0.084  &   0.408   &   0.884     &  0.829 \\
(expt)             &  (0.065)\footnotemark[1]  &   (0.4)     &   (0.91)   &  (0.81)\footnotemark[1] \\
\hline
$\Delta^{\prime}_{{\rm SO}}$ &  0.140  &   0.133   &   0.110     &  0.305 \\
\hline
$\Delta^{-}$ & +0.16 & -0.15 & -0.44 & -0.28 \\
\hline
$m^{\Gamma}_{c}$   &  0.186  &    0.131  &   0.113     & 0.100  \\
(expt)          &  (0.184)  &    (0.13)   &   (0.13)      & (0.096)\footnotemark[2] \\
\hline
$\gamma$      & -0.48  &   +1.29  & +13.3      & +8.5 \\
{}\{Ref.~\onlinecite{Cardona}\}&         &   \{+1.6\}    &             & \{+11.7\}  \\
\end{tabular}
\end{ruledtabular}
\footnotetext[1]{From Ref.~\cite{Landolt}}
\footnotetext[2]{From Ref.~\cite{Kanazawa}}
\end{table}

\begin{figure}[b]
\includegraphics[]{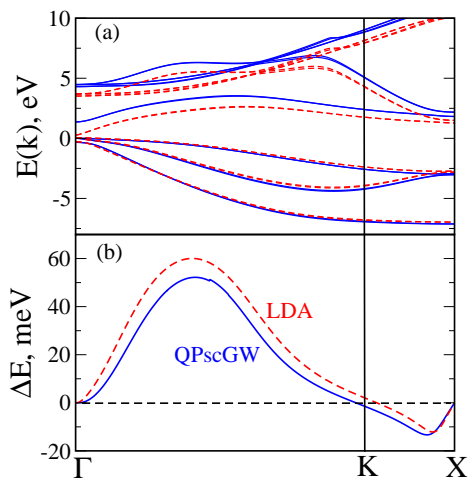}
\caption{\label{fig:gamfbz} (a) The LDA (red dashed line) and QPsc$GW$ (solid blue line) upper valence and lower conduction bands. (b)  The LDA (red dashed line) and QPsc$GW$ (solid blue line) conduction bands spin splitting in [110] direction.}
\end{figure}

In Fig.~\ref{fig:gamfbz} we show the LDA and QPsc$GW$ spin splitting
together with the corresponding bands for GaAs.  Roughly speaking, the spin
splitting of the lowest conduction band is induced by hybridization with
the cation-like conduction band and the anion-like valence bands of $p$
symmetry.  These two bands contribute to $\gamma$ with opposite sign.
Dispersion in the relevant bands on the [110] line differs significantly in
the LDA from the QPsc$GW$ case; but the relative difference is largest near
$\Gamma$ point where the LDA overestimates the $sp$ hybridization.
Consequently the relative difference in LDA and QPsc$GW$ conduction
band spin splitting is largest near $\Gamma$: $\gamma^{LDA}$ is
+121~eV$\cdot\rm{}\AA^{3}$ in GaAs, which is very different from the 6.4 or
8.5~eV$\cdot\rm{}\AA^{3}$ in Table \ref{tab:table1}.  Near X where the
splitting is linear in $k$, LDA and QPsc$GW$ splittings are very similar.
This trend is common to all materials tested.

Until now, several authors have relied on the \kdp\ approximation to
estimate the value of $\gamma$ \cite{Pikus,Cardona,Zawadzki}.  Generally
speaking, the ``downfolded'' \kdp\ Hamiltonian (i.e. with the far band
contributions folded in) can contain all possible independent parameters
consistent with the crystal symmetry at the $\Gamma$ point.  However, it is
not possible to determine these parameters just from experiments.  Thus it
is necessary to assume a number of constraints so as to reduce the number
of the parameters.  For example, a usual approximation is that the
Luttinger's ``residual $\gamma$ parameters'' for the unoccupied $p$ bands
are taken to be zero \cite{Zawadzki}.  However, constraints vary from paper
to paper \cite{Zawadzki,Cardona}, and the spin splitting one obtains
depends on the choice of constraint.  In principle, all parameters can be
calculated from the $GW$ hamiltonian without making any additional
assumptions, in which case the down-folded Hamiltonian should reproduce all
the $GW$ QP levels exactly near $\Gamma$.  However, the parametrization of
\kdp\ theory is based on the Luttinger-Kohn expansion of the
$\psi_{{\bf{k}}n}$ \cite{LK}, where the periodic part of $\psi_{{\bf{k}}n}$
at ${\bf k}\ne{}0$ is expanded by $\psi_{0n}$.  This is neither convenient
nor suitable for the parametrization of \emph{ab initio} eigenfunctions.
It is on the other hand rather straightforward to expand all the required
matrix elements in the basis of the \emph{ab initio} $\psi_{{\bf{k}}n}$
that respect the crystal symmetry.

In conclusion, we employed a recently developed QPsc$GW$ to make \emph{ab
initio} predictions of the conduction band spin splitting in III-V and
II-VI semiconductors.  (In a few cases, a small empirical adjustment to the
self-energy was also needed.)  Such an approach is essential to reliably
determine the splitting; moreover it builds a framework that will enable
builders of model hamiltonians to reliably incorporate the effect without
the need for additional assumptions.


\begin{acknowledgments}

The authors gratefully acknowledge support from the Office of Naval
Research.  A.~N.~Chantis is thankful to X.  Cartoix\'{a} and D.  Z.-Y.  Ting
for fruitful scientific discussions.

\end{acknowledgments}

\bibliography{ggwprl}

\end{document}